\documentclass[aps,reprint,amsmath,amssymb,floatfix,citeautoscript]{revtex4-1}
\usepackage{graphicx}
\usepackage{bm}
\usepackage{braket}
\usepackage{color}
\definecolor{myblue}{rgb}{0,0,1}
\usepackage[breaklinks=true,colorlinks=true,linkcolor=myblue,urlcolor=myblue,citecolor=myblue]{hyperref}

\newcommand{\vk}{{\bm{k}}}
\newcommand{\vp}{{\bm{p}}}

\newcommand{\varRes}{{241}}                         % N
\newcommand{\varTrunc}{{0.10}}                      % k_0 in units of (2 pi/a)
\newcommand{\varExctns}{{80}}                       % M
\newcommand{\varVecPot}{{0.03~\text{a.u.}}}         % (net) A_0
\newcommand{\varVecPotHalf}{{0.015~\text{a.u.}}}
\newcommand{\varVecPotDouble}{{0.06~\text{a.u.}}}   % (net) A_0 (half value)
\newcommand{\varBndGap}{{1.66~\text{eV}}}           % E_g
\newcommand{\varSpinCon}{{148~\text{meV} }}         % lambda_c
\newcommand{\varSpinVal}{{-3~\text{meV} }}          % lambda_v
\newcommand{\varLatCnst}{{3.19~\text{\r{A}}}}       % a
\newcommand{\varHopping}{{1.1~\text{eV}}}           % t
\newcommand{\varPol}{{6.6~\text{\r{A}}}}            % chi_2D
\newcommand{\varBroad}{{10~\text{meV}}}             % line broadening
\newcommand{\varRabSpl}{{67~\text{meV}}}            % Rabi splitting in meV
\newcommand{\varRolMea}{{10~ps}}                    % Transients rolling mean

\begin{document}

\title{Microscopic theory of cavity-confined monolayer semiconductors:\\polariton-induced valley relaxation and the prospect of enhancing and controlling\\the valley pseudospin by chiral strong coupling}

\author{Andrew Salij}

\author{Roel Tempelaar}
\email{roel.tempelaar@northwestern.edu}

\affiliation{Department of Chemistry, Northwestern University, 2145 Sheridan Road, Evanston, Illinois 60208, USA}

\begin{abstract}
We apply a microscopic theory of exciton-polaritons in cavity-confined monolayer transition-metal dichalcogenides including both optical polarizations in the monolayer plane, allowing to describe how chiral cavity photons interact with the valley degrees of freedom of the active material. Upon polariton formation, the degenerate excitons inhabiting the two inequivalent valleys are shown to assume bonding and antibonding superpositions as a result of cavity-mediated intravalley interactions combined with intervalley Coulomb interactions. This is representative of a polariton-induced coherent mixing of the valley polarization. In combination with disorder, this mixing is prone to open a new valley relaxation channel which attains significance with increasing cavity coupling. Importantly, we show that optical cavities with an asymmetric reflectance of left- and right-handed circularly-polarized photons offer a considerably more robust platform to realize a conserved valley polarization, as the valley localization of excitons is reinstated by an asymmetric Rabi splitting which lifts their degeneracy. Moreover, we show this degeneracy lifting to allow for wavelength-selective access to the valley pseudospin by means of a polariton-induced chiral Stark effect, offering interesting opportunities for valleytronic applications.
\end{abstract}

\maketitle

\textbf{Introduction} -- The experimental realization of the strong coupling regime of cavity quantum electrodynamics (QED) offers thrilling prospects for the engineering of functional materials \cite{baranov2018novel}, the control of chemical reactions \cite{ribeiro2018polariton, hertzog2019strong, herrera2020molecular}, and the exploration of new condensed-phase phenomena \cite{keeling2020bose} using photonic degrees of freedom as tuning parameters complementing electronic and nuclear couplings. An even more comprehensive control can be achieved through the deliberate manipulation of spin, charge, and valley quantum numbers. In exploring these possibilities, monolayer transition-metal dichalcogenides (TMDs) are of particular interest \cite{lamountain2018environmental, ardizzone2019emerging, hu2020recent}. TMDs are atomically-thin, direct-bandgap semiconductors \cite{mak2010atomically, splendiani2010emerging} in which charges experience weak dielectric screening due to a reduced dimensionality. This results in strongly-bound electron-hole pairs \cite{chernikov2014exciton, zhu2015exciton} with exceptionally large optical transition dipole moments aligned in the monolayer plane. Such excitons, when hybridized with photonic modes inside optical cavities, yield exciton-polaritons that are stable at room temperature, which combined with their high oscillator strength render TMDs promising candidates \cite{wen2017room} to realize cavity QED based on single quantum emitters \cite{baranov2018novel}.

In-plane inversion symmetry breaking in TMDs yields two inequivalent valleys centered at the corners of the hexagonal Brillouin zone where the bandgaps are located \cite{gunawan2006valley, xiao2007valley}, referred to as $K$ and $K'$. In addition, sizeable spin-orbit coupling \cite{zhu2011giant} results in two non-degenerate optical transitions for each valley, referred to as $A$ and $B$, and interlocks the spin and valley degrees of freedom. Moreover, circularly-polarized light has been predicted \cite{xiao2012coupled} and experimentally demonstrated \cite{cao2012valley, kioseoglou2012valley, mak2012control, sallen2012robust, zeng2012valley} to couple selectively to the excitons inhabiting the $K$ and $K'$ valleys, so that the valley pseudospin can be optically addressed by a simultaneous control of the wavelength and polarization of the incoming light. These properties offer exciting opportunities for the application of TMDs in valleytronic applications, by employing valley polarization or valley coherence. Nevertheless, experimental realizations of valley polarization \cite{cao2012valley, kioseoglou2012valley, mak2012control, sallen2012robust, zeng2012valley, zhu2014exciton, lagarde2014carrier, mai2014many, wang2014valley} and coherence \cite{jones2013optical, hao2016direct, wang2016control} have suffered from high relaxation rates attributed \cite{yu2014valley} to a Maialle--Silva--Sham (MSS) mechanism \cite{maialle1993exciton} involving long-range electron-hole exchange mediated by disorder-induced exciton scattering.

The first observations of strong coupling in cavity-confined TMDs were reported a few years ago \cite{liu2015strong, dufferwiel2015exciton}, in the form of Rabi splitting of optical transitions into polariton branches, and were followed by an increased interest in such systems \cite{schwarz2014two, flatten2016room, lundt2016room, sidler2017fermi, schneider2018two}. A few notable works employing circularly-polarized photoluminescence spectroscopy have reported a retention of the valley polarization of the resulting exciton-polaritons compared to excitons in bare (uncoupled) TMDs \cite{dufferwiel2017valley, sun2017optical, lundt2017observation, chen2017valley}, which was attributed to a weakening of the effects of disorder under strong coupling. A similar improvement of the degree of valley coherence upon polariton formation has been found \cite{dufferwiel2018valley}. These results are suggestive of a viable route towards realizing robust valleytronic behavior, although great strides are to be made in further reducing valley relaxation. A significant hurdle in this endeavor is that theoretical studies on exciton-polaritons in TMDs only began to appear quite recently \cite{karzig2015topological, gutierrez2018polariton, latini2019cavity, chirolli2020brightening}. As such, much remains to be learned about such states, and in particular about the fundamental interaction between cavity modes and the valley pseudospin.

Here, we apply a microscopic model to study the interaction between the excitonic states of TMDs and circularly-polarized cavity modes. Our model is based on the Bethe-Salpeter equation \cite{salpeter1951relativistic}, representing the band structure by a parametrized Dirac-like Hamiltonian \cite{xiao2012coupled}, and generalized to include QED interactions involving both mode polarizations in the monolayer plane. Following this approach we show that cavity-mediated interactions between $A$ and $B$ transitions within the same valley, combined with Coulomb interactions involving valley-differing $A$ and $B$ transitions, coherently mix the degenerate excitons from different valleys, which appear as bonding and antibonding superpositions in the resulting exciton-polaritons. As a consequence, excitation by circularly-polarized light is shown to induce a coherent population transfer between the $K$ and $K'$ valleys. The combined effect of this coherent transfer with disorder is the introduction of a \textit{new} valley relaxation channel that attains prominence with increasing cavity coupling. We propose cavities with an asymmetry in the reflection of left- and right-handed circularly-polarized photons as a design principle for reaching a considerably more robust valley polarization, as it breaks the excitonic degeneracies by valley-selective Rabi splitting. Moreover, the lifted degeneracies grant access to the valley pseudospin based entirely on wavelength-selectivity, by means of a polariton-induced chiral Stark effect.

\begin{figure}[h!]
% Figure is scaled by 50% - to be fixed later
    \centering
    \includegraphics[scale=.5]{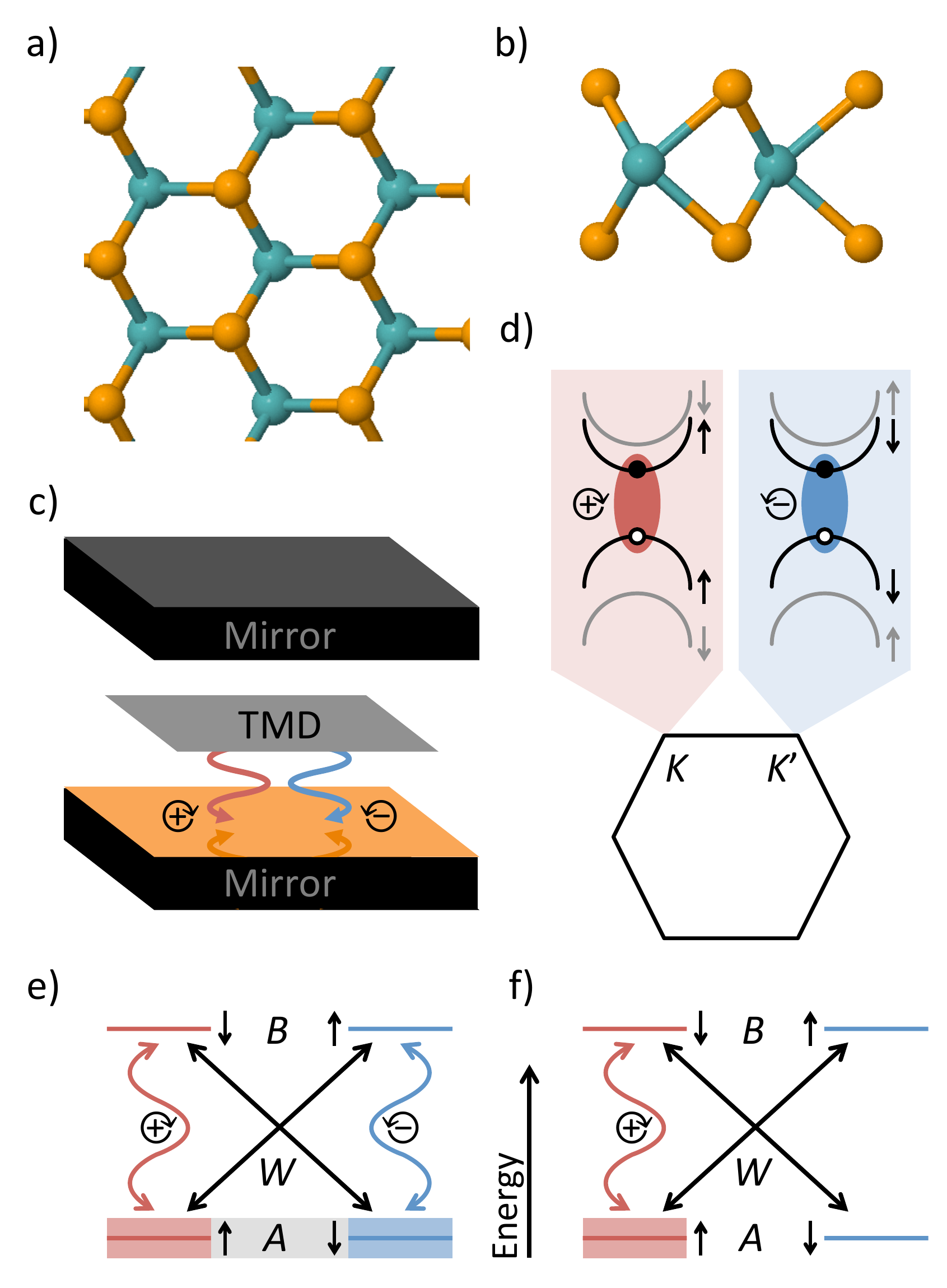}
    \caption{(a,b) Lattice structure of monolayer transition-metal dichalcogenides (TMDs) with transition metals in teal and chalcogens in yellow. Shown are the top view (a) and side view (b) of a monolayer. (c) Schematic of a TMD embedded in an optical cavity. (d) Hexagonal Brillouin zone with $K$ and $K'$ marking the two inequivalent corners. Shown on top is a schematic of the associated spin-dependent valence and conduction band extrema, including the interaction between excitons (electron-hole pairs) and circularly-polarized optical fields. (e) Schematic of the polariton-induced valley relaxation channel. Shown is a simplified energy level diagram of a TMD involving a single excitonic state per $A$ and $B$ transition for each of the $K$ (red) and $K'$ (blue) valleys. Straight arrows represent intervalley Coulomb interactions ($W$), whereas curly arrows denote (effective) intravalley interactions mediated by chiral cavity photons. The latter involves photon emission by the $A$ transition and subsequent absorption by the $B$ transition (and vice versa). The TMD vacuum state, which is occupied in the interim, is not shown. Rabi splittings of the $A$ transition due to resonant cavity modes are depicted as colored shades. The Rabi split levels are fully resonant, promoting a delocalization of the resulting exciton-polaritons over both valleys (grey shade). (f) Same as (e) but for a fully-chiral cavity. Rabi splitting only occurs in the $K$ valley, as a result of which delocalization of the exciton-polaritons is inhibited.}
    \label{fig_scheme}
\end{figure}

\textbf{Theory} -- TMDs are generally defined as $MX_2$ where the transition metal $M$ is molybdenum (Mo) or tungsten (W), and the chalcogen $X$ is sulphur (S), selenium (Se), or tellurium (Te). Figs.~\ref{fig_scheme} (a)-(d) show the lattice structure, and a schematic of the optical couplings of TMDs embedded in a typical Fabry-P\'erot cavity consisting of two parallel mirrors. Optical excitation creates equal-spin electron-hole pairs in either the $K$ or $K'$ valley, resulting in excitons of the form
\begin{align}
\ket{\Phi^m}=\sum_{cv\vk}A_{cv\vk}^md_{c\vk}^\dagger d_{v\vk}\ket{0}.
\end{align}
Here, $m$ labels the excitonic state, $c$ and $v$ run over conduction and valence bands (including spin), $d^{(\dagger)}_{i\vk}$ is the annihilation (creation) operator for an electron with momentum $\vk$ in band $i$, and $0$ is the vacuum level with all conduction bands empty and all valence bands occupied. The excitonic states $\Phi^m$ are obtained by solving the Bethe-Salpeter equation \cite{salpeter1951relativistic}, which is the eigenvalue equation corresponding to the momentum-space electronic Hamiltonian
\begin{align}
H_\text{el}=\sum_{cv\vk}(E_{c\vk}-E_{v\vk})d_{c\vk}^\dagger d_{v\vk}d_{v\vk}^\dagger d_{c\vk}&\\
+\frac{1}{A_\text{2D}}\sum_{cv\vk}\sum_{c'v'\vk'}
\bra{\psi_{c\vk}\psi_{v\vk}^\dagger}K^\text{int}&
\ket{\psi_{c'\vk'}\psi_{v'\vk'}^\dagger}\nonumber\\
&\times d_{c\vk}^\dagger d_{v\vk}d_{v'\vk'}^\dagger d_{c'\vk'}.\nonumber
\end{align}
Here, $E_{i\vk}$ is the energy of an electron with momentum $\vk$ in band $i$, $\psi_{i\vk}$ is the associated (Bloch-type) wavefunction, $A_\text{2D}$ is the area of the monolayer, and $K^\text{int}$ is the electron-hole interaction operator. Note that the daggers appearing in the interaction kernel refer to hole states, being Hermitian conjugates of electron states.

The formation of exciton-polaritons in cavity-confined TMDs is described by supplementing $H_\text{el}$ with the QED terms that nonperturbatively account for electron-photon interactions. Similarly to Refs.~\citenum{gutierrez2018polariton} and \citenum{chirolli2020brightening} we take the long-wavelength limit while neglecting the dipole self-energy term as well as the effects of cavity overtone modes, while adapting a circularly-polarized representation for the cavity field. The latter simplifies our analysis of the chiral properties of polaritons, notwithstanding that a formulation in terms of the more conventional $xy$ representation would yield identical results for cases where cavity modes are symmetric with respect to $x$ and $y$. The resulting Hamiltonian reads
\begin{align}
\label{eq_H_QED}
H_\text{QED}=H_\text{el}+\sum_{\sigma=\pm}\Omega_\sigma a_\sigma^\dagger a_\sigma+\sum_{\sigma=\pm}A_{0\sigma}(P_\sigma a_\sigma^\dagger+\text{H.c.}),
\end{align}
where $a_\sigma^{(\dagger)}$ is the annihilation (creation) operator for a cavity photon with polarization $\sigma$, with $\sigma=+(-)$ referring to right-handed (left-handed) circular polarization in the plane of the monolayer. The associated mode energy $\Omega_\sigma$ is related to the distance between the cavity mirrors, whereas the vector potential amplitude $A_{0\sigma}$ is related to the effective in-plane area of the mirrors. The former can be regarded as independent of the polarization, $\Omega_\sigma=\Omega$, whereas the latter relates to the reflective properties of the mirrors. The electron-photon interaction terms are given by
\begin{align}
    \label{eq_exc_couplings}
    P_\sigma=\sum_{cv\vk}\bra{\psi_{v\vk}}p_\sigma(\vk)\ket{\psi_{c\vk}}d_{v\vk}^\dagger d_{c\vk}.
\end{align}
The circularly-polarized momentum operator elements are related to the $x$ and $y$ polarized analogs as $p_{\pm}(\vk)=(p_x(\vk)\pm ip_y(\vk))/\sqrt{2}$. Solving the eigenvalue equation of $H_\text{QED}$ yields exciton-polariton states of the form
\begin{align}
\ket{\Psi_\alpha}=B_\alpha^+\ket{0,+}+B_\alpha^-\ket{0,-}+\sum_m B^m_\alpha\ket{\Phi^m,0}.
\end{align}
Here, $\alpha$ labels the eigenstate, which is expanded in joint excitonic (first ket index) and photonic (second ket index) contributions. For the latter, $0$ represents the vacuum field of the cavity.

Previous studies have shown the band structure of TMDs in the vicinity of the $K$ and $K'$ points to be well approximated by a parametrized (massive) Dirac-like Hamiltonian given by \cite{xiao2012coupled, ochoa2013spin, cazalilla2014quantum, berkelbach2015bright}
\begin{align}
H(\bm k)=
\left(\begin{array}{cc} -E_\text{g}/2+\lambda_\text{v}\tau_{\bm k}S_z & at(\tau_{\bm k}\bar k_x+i\bar k_y) \\ at(\tau_{\bm k}\bar k_x-i\bar k_y) & E_\text{g}/2+\lambda_\text{c}\tau_{\bm k}S_z \end{array}\right).
\label{eq_twoband}
\end{align}
Here, $a$ is the lattice constant, $t$ is the effective hopping integral, and $E_g$ is the bandgap, which is added with contributions from spin-orbit coupling, quantified by the conduction (valence) band splitting $2\lambda_\text{c(v)}$. $S_z$ is the Pauli spin operator and $\tau_{\bm k}$ is the valley index associated with momentum $\vk$, yielding $+1$ for $K$ and $-1$ for $K'$ \footnote{Different conventions for the valley index, labels, and optical helicity can be found in the literature, e.g., Refs.~\citenum{xiao2012coupled}, \citenum{cao2012valley}, and \citenum{berkelbach2015bright}. Here, we define the  $K$ and $K'$ valleys as those which interact selectively with $+$ polarized and $-$ polarized light, respectively.}. The momentum elements $\bar k_x$ and $\bar k_y$ are capped with a bar to indicate that they appear as the \textit{relative} reciprocal space distance to their nearest Brillouin zone corner ($K$ or $K'$). The Dirac-like Hamiltonian is block-diagonal with respect to the spin index, and for each index value (up and down) yields two bands, as schematically illustrated in Fig.~\ref{fig_scheme} (d). The low- and high-energy bands within each spin block are associated with the conduction and valence band energies $E_{c\vk}$ and $E_{v\vk}$ and corresponding wavefunctions $\psi_{c\vk}$ and $\psi_{v\vk}$, respectively. A known feature of TMDs, the bandgaps at the $K$ and $K'$ points differ considerably for a given spin configuration, with the lower and higher gap energies referred to as the $A$ and $B$ transitions. The momentum operator appearing in Eq.~\ref{eq_exc_couplings} follows from the Dirac-like Hamiltonian as $\vp (\vk)=\frac{m}{\hbar}\nabla_\vk H(\vk)$, where $m$ is the electron rest mass.

Neglecting the exchange interaction between conduction and valence band electrons, the electron-hole interaction kernel is approximated as \cite{berkelbach2015bright}
\begin{align}
\bra{\psi_{c\vk}\psi_{v\vk}^\dagger}K^\text{int}&
\ket{\psi_{c'\vk'}\psi_{v'\vk'}^\dagger}\\
&\approx-\braket{\psi_{c\vk}|\psi_{c'\vk'}}\braket{\psi_{v'\vk'}|\psi_{v\vk}}W(\vk-\vk').\nonumber
\end{align}
In the monolayer limit, the screened Coulomb interaction term is given by the Rytova--Keldysh potential
\begin{align}
W(\vk)=\frac{2\pi e^2}{k(1+2\pi\chi_\text{2D}k)}
\end{align}
where $\chi_\text{2D}$ is the two-dimensional polarizability.

In our calculations, the hexagonal Brillouin zone was discretized using an $N\times N$ Monkhorst--Pack grid (see Supporting Information for details). Since an increasing resolution $N$ effectively corresponds to an increasing monolayer area $A_\text{2D}$, the effective light-matter interactions strength will increase concomitantly. For the sake of our theoretical analysis, it is therefore useful to introduce a net vector potential in which the resolution is divided out, $A_{0\sigma}'=A_{0\sigma}/N$. In order to keep the computational cost manageable while optimizing the sampling accuracy, we have applied a $k$-space truncation radius $k_0$ around the $K$ and $K'$ points \cite{qiu2016screening, tempelaar2019many}.

Without loss of generality, we specifically consider MoS$_2$ as a prototypical TMD, setting $E_\text{g}=\varBndGap$, $\lambda_\text{c}=\varSpinCon$, $\lambda_\text{v}=\varSpinVal$, $a=\varLatCnst$, $t=\varHopping$, and $\chi_\text{2D}=\varPol$.  All reported results were obtained using $N=\varRes$ \footnote{We have deliberately applied an odd resolution, as it better describes the symmetry properties of the excitonic wavefunctions around the $K$ and $K'$ points.} and $k_0=\varTrunc$ (in
units of the inverse lattice constant, $2\pi/a$). Upon evaluating the electronic Hamiltonian, the resulting eigenvalues were shifted by a constant offset such that the lowest-energy eigenstate resides at 2.0~eV. Furthermore, in evaluating $H_\text{QED}$ the joint excitonic-photonic basis set was limited to the first \varExctns~lowest-energy excitons. As shown in the Supporting Information, varying these convergence parameters by a factor of 2 did not yield appreciable differences for the exciton-polaritonic states below 2.2~eV.

\begin{figure}
    \centering
    \includegraphics[]{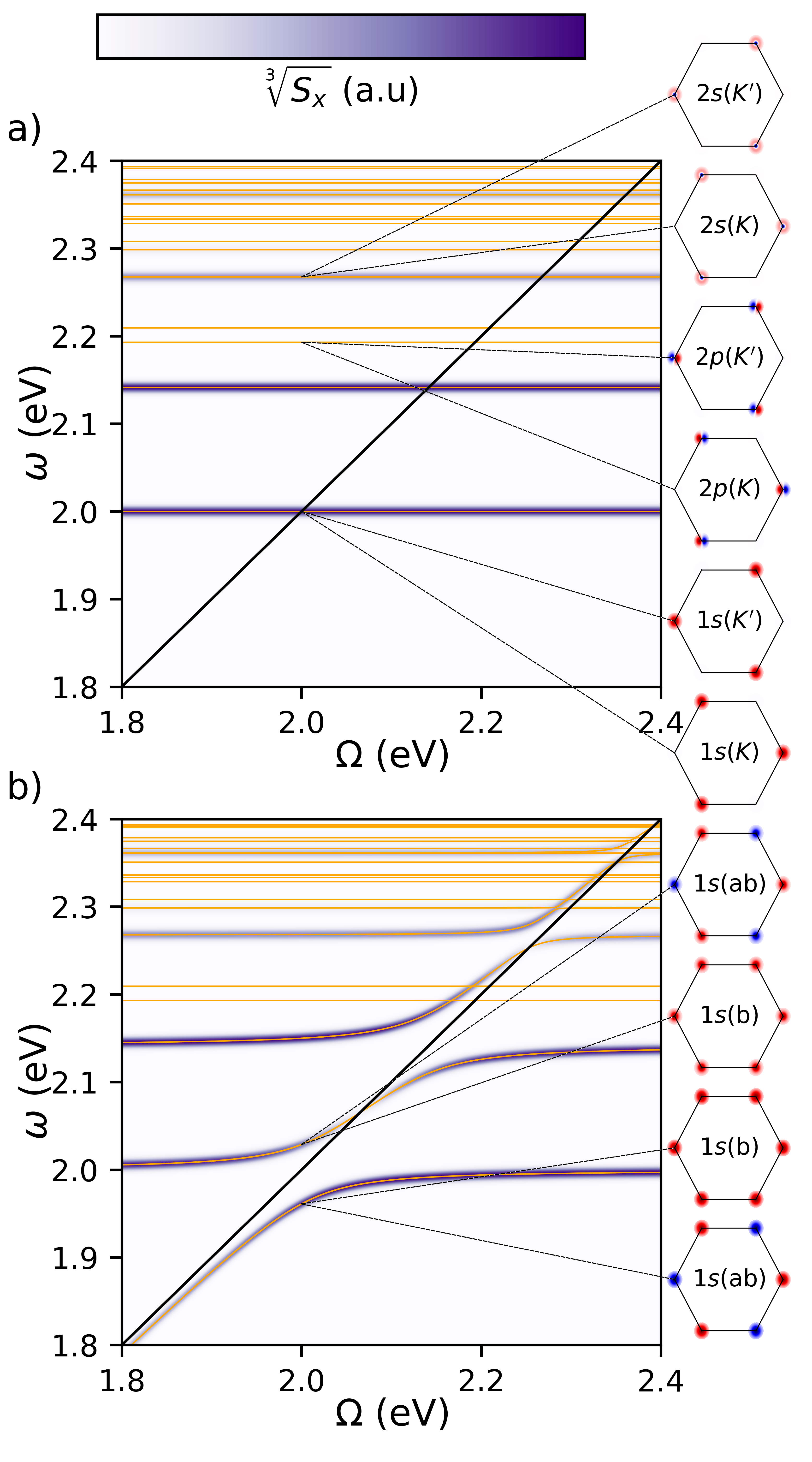}
    \caption{Linearly-polarized optical response $S_x(\omega)$ with varying mode energy $\Omega$ represented as a heat map (with nonlinear scaling to highlight weaker signals). Shown are results for bare MoS$_2$ (a) and for MoS$_2$ under strong cavity coupling with $A_{0\sigma}'=\varVecPot$ (b). Exciton-polariton dispersion relations are indicated with yellow curves. Depicted on the right as heat maps (red is positive, blue is negative) are real values of the excitonic wavefunction envelopes (see text) for select exciton-polaritons, highlighting the $K$ / $K'$ valley localization, or bonding (b) / antibonding (ab) combination.}
    \label{fig_linear_cavity}
\end{figure}

\textbf{Exciton-polaritons} -- Fig.~\ref{fig_linear_cavity} (a,b) compares the linearly-polarized optical response of bare MoS$_2$ (a) and for MoS$_2$ under strong cavity coupling with $A_{0\sigma}'=\varVecPot$ (b). The latter yields a Rabi splitting of the lowest-energy optical transitions of \varRabSpl, in favorable comparison to experimentally reported values \cite{liu2015strong, flatten2016room}. The optical response was calculated using the ``golden rule'' expression
\begin{align}
    S_x(\omega)=\frac{2\pi}{\hbar}\frac{e\tilde{A}_0}{mc}\sum_\alpha\vert\braket{0,0|P_x|\Psi_\alpha}\vert^2F(\omega-\omega_\alpha)
    \label{eq_optical_response}
\end{align}
where $\omega_\alpha$ is the eigenenergy associated with $\Psi_\alpha$, $\tilde{A}_0$ is the vector potential of the incident radiation (not to be confused with the cavity mode vector potentials), $c$ is the speed of light, and $F(\omega)$ is the line shape function, taken to be a Gaussian with a standard deviation of \varBroad. Note that in Eq.~\ref{eq_optical_response} the radiation polarization direction is arbitrarily chosen to align with the $x$ direction, notwithstanding that an extension of our analysis of the optical response to the $y$ direction is trivial.

As seen in Fig.~\ref{fig_linear_cavity} (a), in the absence of cavity coupling the optical response is dispersionless with respect to the mode energy $\Omega$, as expected. A variety of optical transitions are observable, attributable to the nonhydrogenic Rydberg series previously observed for TMDs \cite{chernikov2014exciton} for each of the $A$ and $B$ transitions. Under strong coupling these optical transitions show anticrossing behavior when in resonance with $\Omega$, resulting in significant Rabi splitting in particular for the bright lowest-energy states of the TMD, as seen in Fig.~\ref{fig_linear_cavity} (b). Notably, the exciton-polariton dispersion relations, also shown in Fig.~\ref{fig_linear_cavity} (b), indicate a large number of higher-lying polaritonic states, some of which have been studied in detail in recent works \cite{latini2019cavity, chirolli2020brightening}.

In order to evaluate the anatomy of the excitons and exciton-polaritons shown in Fig.~\ref{fig_linear_cavity} (a,b), it proves convenient to introduce the excitonic wavefunction envelope, defined as
\begin{align}
    C_\alpha(\vk)\equiv\sum_{cv}\bra{\psi_{v\vk}}p_x(\vk)\ket{\psi_{c\vk}}\braket{0,0|d_{v\vk}^\dagger d_{c\vk}|\Psi_\alpha},
\end{align}
which is essentially the $\vk$ resolved optical interaction element appearing in Eq.~\ref{eq_optical_response}. Shown in Fig.~\ref{fig_linear_cavity} (a,b) alongside the optical response and dispersion relations are real values of such envelopes for select excitons and exciton-polaritons at $\Omega=2.0$~eV. (We restrict ourselves to depicting states associated predominantly with $A$ transitions, although noting that those associated with $B$ transitions are very similar in nature, albeit shifted to higher energies.) These results reflect the aforementioned nonhydrogenic Rydberg series, identifying the lowest-energy states with $1s$, while informing on the valley-polarization of the excited states.

From the excitonic wavefunction envelopes shown in Fig.~\ref{fig_linear_cavity} (a,b), it is obvious that while the excited states in bare MoS$_2$ are localized in either the $K$ or the $K'$ valley, those under strong coupling have become delocalized over both valleys. The mechanism responsible for this delocalization is depicted in Fig.~\ref{fig_scheme} (e), and involves a combination of \textit{intervalley} Coulomb interactions ($W$) involving same-spin $A$ and $B$ transitions, and \textit{intravalley} couplings between the same transitions of opposite spin. The latter is mediated by the emission and subsequent absorption of a chiral cavity photon (with the TMD residing in the vacuum state in the interim). Although each $A$-$B$ pair is highly nonresonant, the combination of all four pairwise interactions effectively results in a coupling between fully-resonant $A$ transitions associated with different valleys (through a mechanism akin to superexchange), and similarly for the $B$ transitions. As a result, upon polariton formation we observe the $1s$ excitons of the $K$ and $K'$ valleys to engage in bonding and antibonding superpositions, yielding a pair of bright and dark polaritons for each branch. Similar behavior is observed for all other optically-accessible excitons.

\textbf{Valley relaxation} -- The observed bonding and antibonding superpositions of valley-differing excitons under polariton formation are representative of a coherent mixing of the valley polarization not seen for bare TMDs. This may seem to be at odds with previous experimental reports indicating that strong coupling acts as to conserve valley polarization \cite{dufferwiel2017valley, sun2017optical, lundt2017observation, chen2017valley}. In that regard, it is important to note that our model does not include disorder, which is expected to break the perfect degeneracy between excitons from different valleys, and inhibit the formation of bonding and antibonding superpositions. However, in the presence of disorder, the interactions responsible for this coherent mixing provide the conditions for a \textit{new} valley relaxation channel, acting in concert with the MSS channel. The relaxation rate of this new, polariton-induced channel is generally expected to increase with the effective coherent interactions while decreasing with the degree of disorder. Even though a quantitative analysis of the interplay of coherent interactions and disorder is beyond the scope of the present study, it is instructive to consider the coherent evolution of valley-selective optical excitations under strong coupling in the absence of disorder, as it provides an upper bound for the relaxation rate associated with the polariton channel.

Reported measurements of valley polarization have probed the fate of valley-selective optical excitations by using circularly-polarized photoluminescence spectroscopy with chiral control of the pump pulse, monitoring the anisotropy
\begin{align}
    \eta(t) = \frac{I_+(t)-I_-(t)}{I_+(t) + I_-(t)}.
\end{align}
Here, $I_\pm(t)$ is the $\pm$ polarized signal intensity, with $t$ loosely-defined as the time interval between the pump pulse and the signal detection. Assuming a $+$ polarized pump pulse and impulsive excitation and detection events (corresponding to an infinite frequency bandwidth), this intensity is governed by the equation
\begin{align}
I_\pm(t)\propto\vert\braket{\Psi_\pm(0)|\Psi_+(t)}\vert^2,
\end{align}
where the proportionality constant is omitted as it cancels in the expression for $\eta(t)$. Here, $\ket{\Psi_\pm(t)}$ represents the excited state at a time $t$ after excitation with a $\pm$ polarized pulse,
\begin{align}
    \ket{\Psi_\pm(t)}= e^{-iH_\text{QED}t}P_\pm^\dagger\ket{0}.
\end{align}
Expressions corresponding to a pump with $-$ polarization follow analogously.

\begin{figure}
    \centering
    \includegraphics{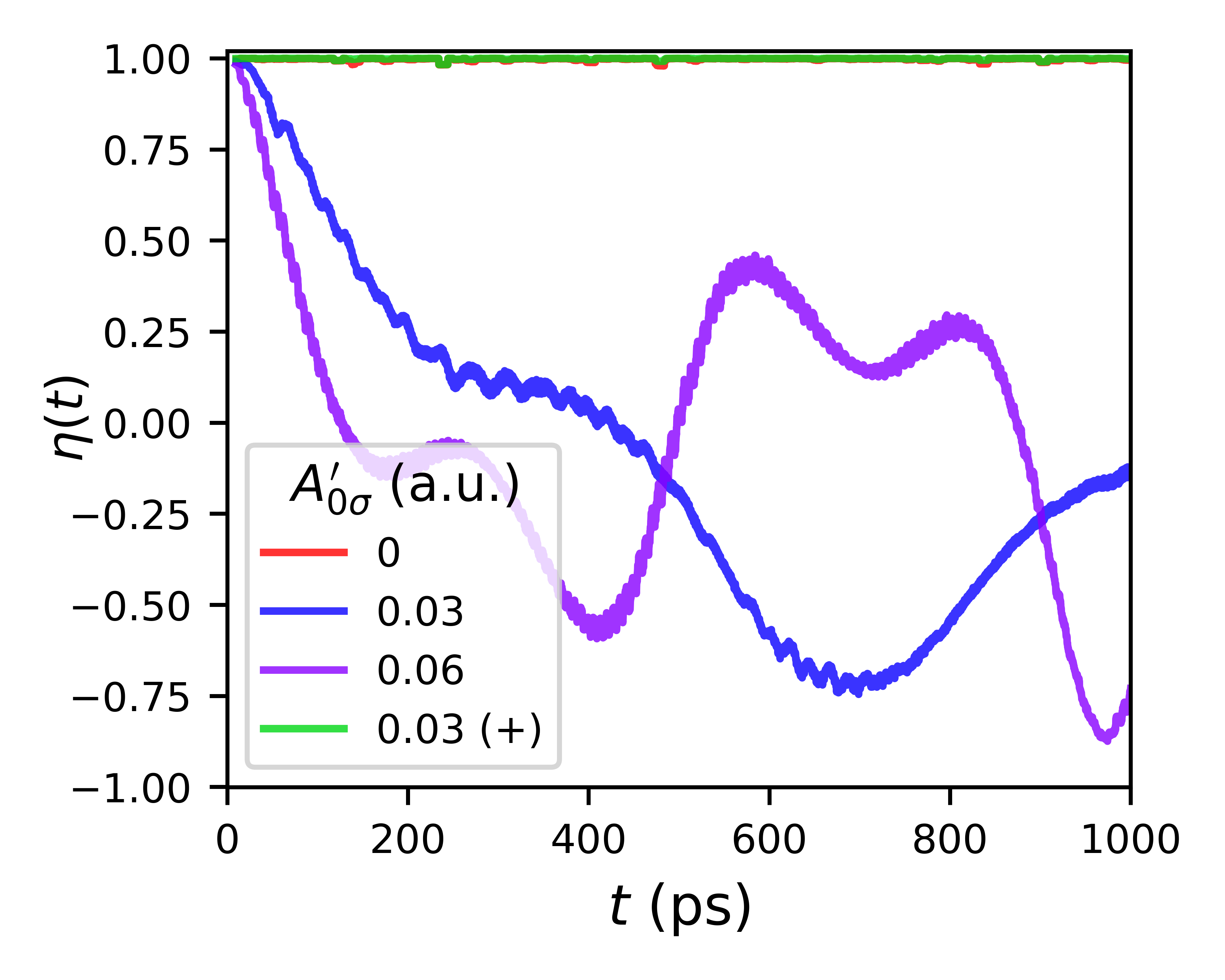}
    \caption{Chiral anisotropy at time $t$ after $+$ polarized optical excitation at various cavity coupling strengths and polarizations. Results have been subjected to a rolling mean of \varRolMea. All shown results are for cavities with $A_{0+}=A_{0-}$, except for the curve labelled ($+$), for which $A_{0-}=0$.}
    \label{fig_anisotropy}
\end{figure}

Shown in Fig.~\ref{fig_anisotropy} is the time-dependent anisotropy $\eta(t)$ of cavity-confined MoS$_2$ at different values of the vector potential $A_{0\sigma}'$, assuming a $+$ polarized pump pulse. The involvement of a multitude of coupled higher-energy excitons resulted in high-amplitude quantum beats in the raw data at frequencies unresolvable under typical experimental conditions. We have therefore applied a rolling mean of \varRolMea~in order to focus on the slower (average) trend. Also shown as a reference is the result for bare MoS$_2$ ($A_{0\sigma}'=0$). Since our model does not include disorder, the MSS channel is inactive, which combined with the absence of cavity coupling yields a perfectly-conserved valley polarization. This is reflected by the near-perfect retainment of the anisotropy, only to be interrupted by intermittent spikes caused by the weak intervalley Coulomb interactions involving spin-identical $A$ and $B$ transitions. In marked contrast, the anisotropy obtained at $A_{0\sigma}'=\varVecPot$ shows significant beatings between valleys resulting from the polariton-induced coherent mixing of valley-differing excitons. Moreover, results obtained for an even stronger coupling of $A_{0\sigma}'=\varVecPotDouble$ exhibit a significant decrease in the beating time. Taking the instant at which the anisotropy first crosses the zero line as a qualitative measure of the valley relaxation time, we observe this time scale to decrease with increasing cavity coupling.

The above results suggest that for a maximally-conserved valley polarization one has to optimize $A_{0\sigma}'$ by minimizing the combined effect of the polariton channel and the MSS channel. This makes for an intriguing scenario wherein strong coupling inhibits the MSS channel while promoting the polariton channel, whereas disorder promotes the MSS channel while inhibiting the polariton channel (through degeneracy breaking). Importantly, in cases where the MSS channel is the dominant factor, increasing cavity coupling only improves the valley polarization by its suppression of this dominant effect. This appears to be borne out in relevant experimental reports \cite{dufferwiel2017valley, sun2017optical, lundt2017observation, chen2017valley}. Interestingly, low-temperature measurements by Chen \textit{et al.} showed bare TMDs to feature a higher degree of valley polarization than their coupled counterparts, while at elevated temperatures this trend was reversed \cite{chen2017valley}. This is consistent with our theoretical findings, considering that disorder is known to generally increase with temperature. As such, the polariton channel may be dominant at low temperatures, as a result of which bare TMDs exhibit a comparatively higher degree of valley polarization, while with increasing temperature the MSS channel becomes dominant, yielding a comparatively higher degree of valley polarization for strongly-coupled TMDs instead. As such, our findings offer a potential explanation of this experimental observation.

\textbf{Chiral cavities} -- In view of the above considerations, intriguing possibilities arise when confining TMDs in cavities that involve an asymmetry in the reflectance of right- and left-handed circularly-polarized light \cite{orr2014design}. In the limit where reflection occurs exclusively for one chirality, the intravalley interactions involving photons of the other chirality are fully eliminated; see Fig.~\ref{fig_scheme} (f). More importantly, only excitons inhabiting the valley with a chirality matching that of the remaining cavity photons will experience strong coupling and exhibit Rabi splitting. This breaks the excitonic degeneracies, which inhibits the formation of bonding and antibonding superpositions. As such, it will act similarly to disorder in suppressing the polariton channel, but without activating the MSS channel, and with a degree that increases with cavity coupling.

\begin{figure}
    \centering
    \includegraphics[]{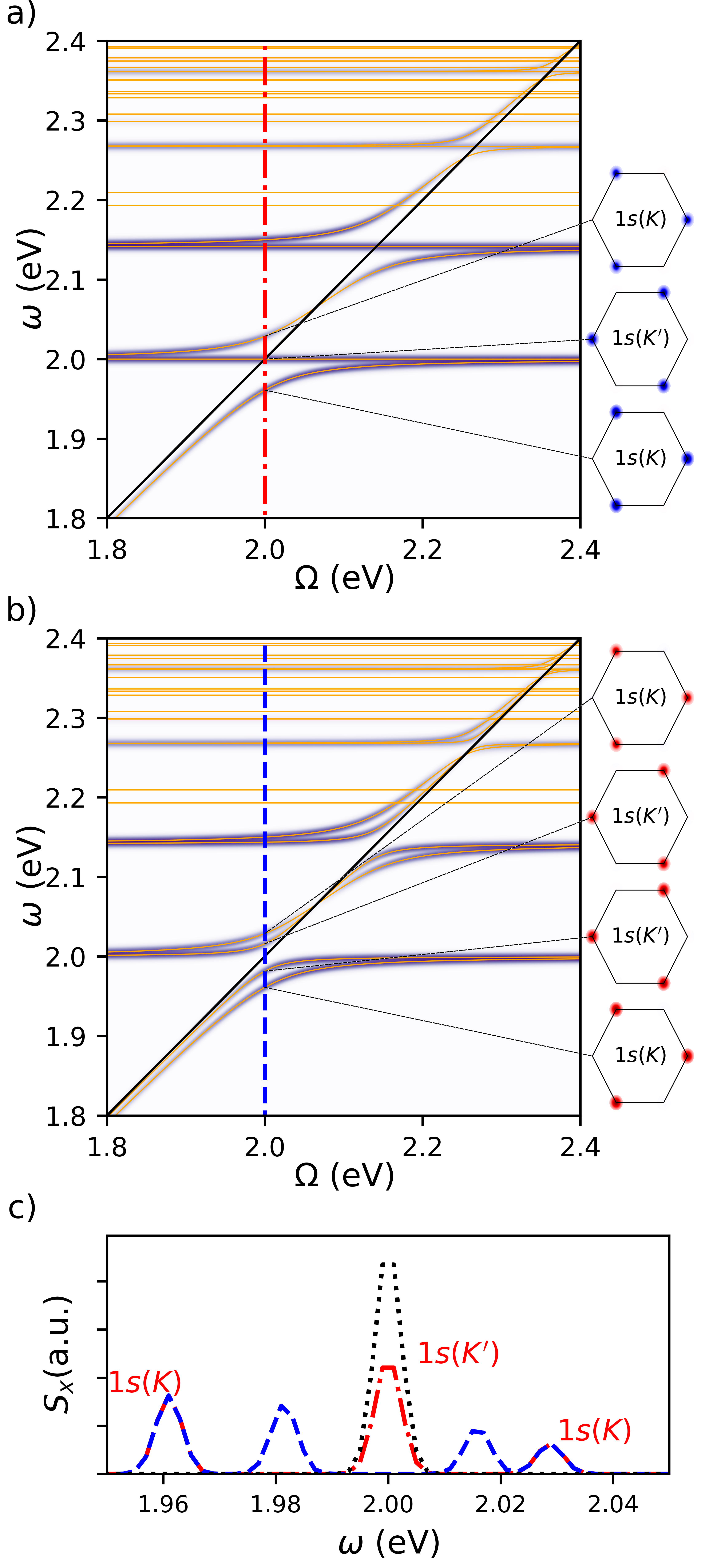}
    \caption{(a,b) Same as in Fig.~\ref{fig_linear_cavity}, but for a chiral cavity with $A_{0+}'=\varVecPot$ and $A_{0-}'=0$ (a), and $A_{0+}'=\varVecPot$ and $A_{0-}'=\varVecPotHalf$   (b). (c) Linearly-polarized optical response at $\Omega=2.0$~eV taken from (a) and (b) as well as from Fig.~\ref{fig_linear_cavity} (a), represented as dot-dashed red, dashed blue, and dotted black, respectively. Valley-localized exciton and exciton-polariton peaks are labelled for the results from (a).}
    \label{fig_chiral_cavity}
\end{figure}

The above predictions are borne out in Fig.~\ref{fig_chiral_cavity} (a,b), where results are shown for MoS$_2$ optically-confined in such chiral cavities. Fixing $A_{0+}'=\varVecPot$, a fully-chiral cavity was parametrized as $A_{0-}'=0$ whereas a partially-chiral analogue was parametrized as $A_{0-}'=\varVecPotHalf$. The linearly-polarized optical response of the fully-chiral cavity, shown in Fig.~\ref{fig_chiral_cavity} (a), demonstrates a Rabi splitting for only one of the two degenerate excitonic states, with the other state remaining completely dispersionless. Moreover, the associated excitonic wavefunction envelopes show that both excitons and exciton-polaritons have reassumed their valley-localized character, and under the applied + polarized cavity, Rabi splitting occurs exclusively for polaritons involving $K$ valley excitons. For the partially-chiral cavity, shown in Fig.~\ref{fig_chiral_cavity} (b), the excitons are seen to organize in two polariton pairs, but with unequal Rabi splittings. Excitonic wavefunction envelopes are found to be fully valley-localized also in this case. The above results indicate valley polarization to be conserved under chiral cavity coupling. This is further substantiated by the time-dependent anisotropy shown for the fully-chiral cavity in Fig.~\ref{fig_anisotropy}. When compared to the non-chiral cavity results for the same vector potential, this data shows chiral cavity confinement to be highly effective in inhibiting valley relaxation. Moreover, since the Rabi-split pairs are otherwise behaving as regular exciton-polaritons, it is expected that the suppression of the MSS channel is maintained under these conditions. Hence, with a suppression of both polariton and MSS channels, we expect chiral cavities to hold great potential for the realization of robust valley polarization.

Another interesting possibility arises as a result of the asymmetric Rabi splittings found for the $K$ and $K'$ localized excitons under chiral coupling, which breaks the degeneracy found for such excitons in bare TMDs. As a result of this degeneracy breaking, the valley pseudospin can be addressed wavelength-selectively \textit{regardless of the light polarization}. This is akin to the valley-selective Stark effect brought about when applying intense circularly-polarized light to TMDs \cite{sie2015stark}, but without the need for optical pumping.

The polariton-induced valley-selective Stark effect is represented more clearly in Fig.~\ref{fig_chiral_cavity} (c), showing the linear absorption spectra of MoS$_2$ confined in chiral cavities with $\Omega=2.0$~eV; tuned to be resonant with the $1s$ excitons associated with the $A$ transition. When comparing the spectra of MoS$_2$ confined in a fully-chiral cavity with that of bare MoS$_2$, the $A$ transition is seen to split into three peaks, two of which comprise the upper and lower polariton branch associated with the $K$ valley, and the third unshifted peak associated with the undispersed exciton localized in the $K'$ valley. For the partially-chiral cavity, four peaks are found, constituting two polariton pairs, while the individual peaks still provide wavelength-selective access to the associated valley pseudospin.

\textbf{Conclusions and discussion} -- By applying a microscopic theory of TMDs confined in optical cavities we have shown that strong coupling promotes a coherent mixing of the two valley pseudospin states. Combined with disorder, this is prone to opening a new polariton-induced valley relaxation channel that gains significance with increasing cavity coupling. While polariton formation purportedly improves valley polarization by suppressing the MSS channel, the polariton channel counteracts this beneficial effect. The resulting interplay of disorder and strong coupling bears signatures that are consistent with previous experimental reports \cite{chen2017valley}. Interestingly, under conditions where disorder is minimal, our results suggest strong coupling to provide an opportunity to coherently manipulate the valley pseudospin by its ability to rotate the stationary solutions for the valley polarization.

It has recently been shown theoretically that bonding and antibonding superpositions of excitons, similar to those observed in the present work, may result from a finite width of the optical cavity, which introduces couplings of valley-localized excitons to photons of opposite chirality \cite{gutierrez2018polariton}. These effects are not accounted for in the present study, although it should be noted that the intervalley Coulomb interactions give rise to an effective coupling of the very same general form, as it essentially borrows oscillator strength of opposite chirality to the excitons. It is therefore likely that the mechanism discussed in Ref.~\citenum{gutierrez2018polariton} contributes to the polariton channel discussed in the present study, and that its relative contribution depends on the cavity parameters. It is also noteworthy that at extreme coupling strengths a mixing of bright and dark excitons has been shown to impact the optical response at higher energies \cite{latini2019cavity}. It will be interesting to expand the present model with contributions from the dipole self-energy term and to investigate how these mixing behaviors are complementing the chiral polaritonic effects.

Promising opportunities arise when applying optical cavities with an asymmetrical reflectance of right- and left-handed circularly-polarized light. Our study proposes such a setup as a superior platform to engineer robust valley polarization, as the polariton-induced relaxation channel becomes strongly suppressed. Moreover, the asymmetry in chiral cavity modes yields valley-dependent Rabi splittings, which breaks the degeneracy of the $K$ and $K'$ excitonic transitions by means of a polariton-induced chiral Stark effect. This enables one to address the valley pseudospin wavelength-selectively, irrespective of the light polarization, and holds potential implications for the manipulation of valley coherence \cite{ye2017optical}. It may be interesting to consider the possibility to employ similar effects for cases involving other chiral optical resonators \cite{chervy2018room, tang2019on-chip, wu2019room, wen2020steering, li2020room, kim2020valley} or plasmonic nanogaps \cite{sun2020selectively}. It may also be worthwhile to substitute monolayer TMDs with few-layer analogs or graphene-TMD heterostructures, which have been shown to exhibit an enhanced intrinsic valley polarization \cite{lorchat2018room, godiksen2020contrast}. Such implementations offer exciting prospects for valleytronic control, designing optoelectronic Hall devices \cite{xiao2012coupled, lundt2019optical}, as well as realizing circularly-polarized lasing at room temperature \cite{huang2020trends}.

\textbf{Acknowledgements} -- This work was supported by startup funds from Northwestern University.

\bibliography{bibliography}

\end{document}